# Symmetry-driven atomic rearrangement at a brownmillerite-perovskite interface

*Tricia L. Meyer*[1], *Hyoungjeen Jeen*[1], *Xiang Gao*[1], *Jonathan R. Petrie*[1], *Matthew F. Chisholm*[1], *Ho Nyung Lee*[1]*

[1]Materials Science and Technology Division, Oak Ridge National Laboratory, Oak Ridge, Tennessee 37831, United States

*Email: hnlee@ornl.gov

The design of new materials has transformed considerably over the past decade from an emphasis on tailoring bulk properties to those isolated at the interface between two materials. Undoubtedly, many would argue that the interface is the key to new, multifunctional materials and devices. Examples include the discovery of a 2-dimensional electron gas (2DEG) and thermopower enhancement in $LaTiO_3/SrTiO_3$ (STO) heterostructures[1–3], electric-field control of spin polarization between ferroelectric and ferromagnetic film layers[4] and colossal ionic conductivity at the yttria-stabilized $ZrO_2$ (YSZ)/STO interface[5]. The strong property modification in heterostructures such as these suggests that the structural, electronic, lattice and orbital degrees of freedom at the interface can be tuned from the individual constituents.

Many of the interfacial properties that have drawn significant attention have been linked to structural effects induced by lattice or symmetry mismatch.[5–16] Symmetry mismatch occurs when the interface is formed from two non-isostructural materials such as an orthorhombic film on a cubic substrate, or even materials consisting of different coordination environments. While it is possible for non-isostructural bilayers to contain a negligible *average* lattice mismatch, low symmetry materials grown on higher symmetry substrates will likely exhibit an unavoidable non-uniform strain due to different lattice parameters along the in-plane directions (*e.g.* orthorhombic unit cell where a≠b≠c).[12] Thus, heterostructures containing this type of interface often result in interfacial strain of the cation and anion sublattices leading to atomic reconstruction or disorder and interfacial properties different from the bulk.[5,14,15]

*This manuscript has been authored by UT-Battelle, LLC under Contract No. DE-AC05-00OR22725 with the U.S. Department of Energy. The United States Government retains and the publisher, by accepting the article for publication, acknowledges that the United States Government retains a non-exclusive, paid-up, irrevocable, world-wide license to publish or reproduce the published form of this manuscript, or allow others to do so, for United States Government purposes. The Department of Energy will provide public access to these results of federally sponsored research in accordance with the DOE Public Access Plan (http://energy.gov/downloads/doe-public-access-plan).*

One such example of this type of non-isostructural interface exists between the brownmillerite, $SrCoO_{2.5}$ (SCO), and cubic perovskite, STO ($a_c$ = 3.905 Å). The brownmillerite structure is derived from the perovskite phase ($SrCoO_{3-\delta}$) through the removal of $1/6^{th}$ of the oxygen atoms; thereby reducing the crystal symmetry from the highly symmetric cubic phase (space group $Pm\bar{3}m$) to a lower symmetry orthorhombic phase ($a_o$ = 5.5739(2) Å, $b_o$ = 5.4697(2) Å, $c_o$ = 15.7450(5) Å).[17–20] Interestingly, this structural phase transformation is accompanied by a change in properties from a ferromagnetic metal to an antiferromagnetic insulator as $\delta$ approaches 0.5.[21,22] While the lattice mismatch between SCO and STO is negligible when considering the average in-plane lattice parameter for SCO ($a_{pc}$= 3.905 Å), compressive and tensile strains near 1% are calculated with respect to the individual $a_o$ and $b_o$ lattice parameters, inducing a non-uniform strain of the brownmillerite lattice. Furthermore, the brownmillerite structure consists of alternating $CoO_6$ octahedral and $CoO_4$ tetrahedral layers forming a network of ordered 1-D oxygen vacancy channels, which are shown to facilitate fast catalytic activity and high oxygen mobility.[23,24] Thus, the different coordination environments in SCO provide the potential for two different STO:SCO interfaces, one containing the octahedral layer or one containing the tetrahedral layer as the nucleation layer.

Provided the highly anisotropic oxygen diffusion through the vacancy channels and the complex electronic and magnetic structure of SCO, a thorough investigation of the interfacial nucleation, connectivity and microstructure of this non-isostructural interface could provide unique insight into the growth of layered oxides and their nanoscale phenomena. Thanks to the advancements in imaging techniques over the past several decades, the structural components contributing to these interfacial phenomena can be identified and used in the interpretation of new nanoscale behavior.[25–28] In this work, we probe the microstructure of epitaxial SCO films grown on STO substrates using scanning transmission electron microscopy (STEM), which can reveal the influence of symmetry mismatch-driven strain upon the interfacial architecture. Using geometric phase analysis[29,30] (GPA), we carefully mapped out the strain-variation existing between two different domain structures. We also identified asymmetric cation displacements at the interface that occur in order to accommodate the mismatch of lattice and symmetry presented across the oxygen vacancy channels.

The high crystalline quality of the SCO film is shown by annular dark field (ADF)-STEM images in **Figure 1**. Two domain structures, defined as $a_\parallel$ and $b_\parallel$ domains, are observed in the brownmillerite film when viewed along the [100] and [010] directions, respectively, which are parallel to the STO [110] direction. The ADF column intensities are approximately proportional to the atomic number ($Z$) of each element present to the power of $x$ ($x \sim 2$). Thus, the columns containing strontium are the most intense, the columns containing cobalt atoms are less intense, and the oxygen columns are nearly invisible. A distinct stripe formation representative of an underlying superstructure of the brownmillerite phase can also be seen. The stripes are the oxygen-deficient tetrahedral layers that have a diminished intensity compared to the octahedral layers, mostly due to the expanded [001] (out-of-plane) Sr−Sr distance in these layers of ordered oxygen vacancies, which is consistent with the characteristic structure of other brownmillerite phases.[31] The oxygen vacancy channels are shown in Figure 1b where the brownmillerite structure is oriented along the [010] direction. The clarity of the oxygen vacancy channels further verifies the high quality crystal growth achieved with pulsed laser epitaxy (PLE) of these layered oxides and shows features (e.g. oxygen vacancy channels) that are representative of the characteristics bulk structure of the brownmillerite phase.

While it can be difficult to identify the space group symmetry of brownmillerite SCO[20,32,33], Figure 1 can provide a hint to the possible space group symmetry of our film. Muñoz *et al*. determined from neutron diffraction coupled with theoretical calculations that the *Ima*2 space group, in which the tetrahedral chains maintain the same orientation in the alternating tetrahedral layers, is the most energetically stable .[20] In contrast, Sullivan *et al*. proposed the space group, *Imma*, in which the tetrahedral chains have different orientations in the alternating layers.[33] The question then becomes whether the tetrahedral chains maintain the same orientation or whether they are disordered. We cannot easily discern between these space groups when viewing the structure along the [010] direction (Figure 1b), since this view is similar between the *Imma* and *Ima*2 space groups. However, the 90° rotated domain shown in Figure 1a can provide some hints. Here, the bright contrast of the Co and Sr atoms are quite distinct. If the tetrahedral chains were disordered, we would expect "smearing" of the tetrahedral cobalt atoms since the averaged structure would contain tetrahedrally-coordinated cobalt atoms slightly offset from one another. Thus, the configuration of tetrahedral chains in the *Ima*2 space group may be more representative of our films. Though it is beyond the scope of this work, Raman spectroscopy could reveal further insight into the tetrahedral chain configurations.[34]

Interestingly, the $b_\parallel$ domain contains a distinct contrast variation at the interface, similar to a strain field, which is not visible for the remainder of the film nor when viewing along the [100] orientation. This variation is surprising since such contrast often occurs when the lattice mismatch or dislocation density is large, which is not the case here. We have observed this strain contrast in all images taken along the [010] direction. Furthermore, we performed reciprocal space mapping (not shown) of this film which showed that SCO was coherently strained with the substrate. Thus, the anomalous interfacial phenomenon likely originates from factors other than extrinsic defects.

In order to investigate the anomalous strain field within the $b_{\parallel}$ domain, GPA was employed. Using this technique, we were able to determine atomic displacements and map lattice strain in our high resolution STEM images with respect to a reference structure, i.e. our STO substrate. Since the strain determined from GPA is a relative value, it is important to mention that it can only designate whether the measured lattice is larger or smaller than the reference lattice.[35] We selected an HAADF image (**Figure 2a**) covering both $a_{\parallel}$ and $b_{\parallel}$ domains labeled regions I and II, respectively, and provide a strain map of the image in Figure 2b. Here, it is evident that the magnitude of the in-plane strain, $\varepsilon_{xx}$, is different for the two domain directions. We investigate the relative strain in more detail by extrapolating the values from intensity line-scan profiles obtained from the strain map in Figure 2c. If we assume that the substrate is nominally strain-free ($\varepsilon_{xx} \sim 0$), it is clear that region II contains a larger magnitude of in-plane strain along the [100] direction than along the [010] direction in region I.

While less strain is observed across the oxygen vacancy channels, the presence of a strain field suggests that the interfacial structure is in some way different from the bulk structure. In order to understand the interfacial behavior, the nucleation layer of the cobaltate on the $TiO_2$-terminated substrate must be identified since two different interfaces are possible: (1) $SrCoO_{2.5}$······SrO-CoO-SrO-$\underline{CoO_2}$-$\underline{SrO}$-$TiO_2$-SrO······$SrTiO_3$, where the interfacial termination on the $SrCoO_{2.5}$ side is an octahedral layer or (2) $SrCoO_{2.5}$······SrO-$CoO_2$-SrO-$\underline{CoO}$-$\underline{SrO}$-$TiO_2$-SrO······$SrTiO_3$, where the interfacial termination is a tetrahedral layer. Surprisingly, interfaces such as this have not been explored in detail. This could be due to the logical assumption that the substrates' octahedra may serve as a template for the first layer of the deposited SCO.[8] In an effort to identify the role of symmetry mismatch at the brownmillerite-perovskite interface and determine the nucleation layer, we have collected ADF (**Figure 3a**) and complimentary annular bright field (ABF) (Figure 3b) images of the same region, which provide visuals of the Sr and Co atoms in their lattice positions. More importantly, the oxygen column positions are revealed in the ABF image (white circle) as are the oxygen vacancy columns (red square). We also observe tilting of the $CoO_6$ octahedra that occurs in order to accommodate the $CoO_4$ tetrahedra. Close analysis of the interface region shows that the epitaxial growth of SCO starts with the tetrahedral layer. This observation of the nucleation layer is important when considering interfacial connectivity since the planar symmetry of the tetrahedral layer and $TiO_2$-terminated layer is dissimilar. Specifically, the interface is composed of evenly spaced titanium atoms while the cobalt atoms within the tetrahedral layer have two different atomic spacings that alternate along the length of the interface. As will be discussed in more detail later, the presence of the local contrast variation and the different planar connectivity at the interface is a strong indication that the interfacial sublattice of the tetrahedral layer is displaced. It should also be mentioned that while the interfacial layer is tetrahedral, a light contrast can be seen between the cobalt columns in Figure 3b, indicating that the expected oxygen vacancy channels are not completely empty in some regions. Since the substrate is $TiO_2$-terminated, it is possible that some step terraces at the interface may be overlapping with the film layer, revealing additional contrast along the atomic columns. While it is beyond the scope of this work, we would like to point out that there are still outstanding questions to be further addressed – Is the type of interfacial layer selectively controllable between tetrahedra and octahedra? Can variation of oxygen concentration be a means to control the termination?

In order to better understand the interfacial structure, we analyzed the average in-plane Ti and Co displacements along an atomic column and out-of-plane Sr−Sr interatomic spacing as a function of increasing $N$ in **Figures 4a** and **4b**. The displacements and interatomic spacing were extracted from an HAADF image of the $b_\parallel$ domain by averaging values of neighboring columns for 24 unit cells and 48 unit cells for in-plane and out-of-plane measurements, respectively. We include the atomic spacing of the substrate (highlighted in green) since these are known to influence the first few layers of the film. With an average diameter of 3.61(8) Å, the separation between tetrahedral Co across the channel is nearly 81% wider than the directly adjacent tetrahedra with an average in-plane distance of 2.00(1) Å. Due to the greater dimensions of the tetrahedral layers, it is clear why the activation energy of oxygen migration in SCO determined from density functional theory is lowest through the oxygen vacancy channels.[23,24,36] A similar observation was found for the well-studied oxide ion conducting brownmillerite, $Ba_2In_2O_5$, which verifies the universal nature of anisotropic oxygen diffusion through brownmillerites.[37,38]

The region highlighted in pink illustrates the distinct inhomogeneity of the interfacial atomic spacing that was visible in the strain-field region shown in Figure 1b. Here, the most notable observation is the large, displacement of Co atoms at the $N = 1$ layer to positions reminiscent of octahedrally-coordinated Co, despite their tetrahedral coordination. These displacements are consistent with a systematic expansion and contraction of the spacing between neighboring tetrahedra at the interface. The lack of change in the subsequent octahedral Co−Co distances suggests that the octahedral rotation within these layers is relatively unaffected by the modulated tetrahedral distances. If we consider the out-of-plane Sr−Sr interatomic spacing across the octahedral and tetrahedral layers, we see that they are relatively uniform until the interface region and surface layer of the substrate, which show a slight expansion coinciding with the in-plane displacements. Interestingly, the interfacial reconstruction of the CoO$_4$ sublayer and the surrounding Sr atoms does not alter the rigid TiO$_6$ octahedra on the surface of the substrate. Provided the high uniformity of the film beyond $N = 2$, we reiterate that the structural quality of the film is excellent and our growth control is quite high as evidenced by our previous reports of sharp interfaces.[39] Considering this fact, we believe that the modulation in the atomic architecture at $N = 1$ across the vacancy channels is a way to minimize the interfacial formation energy required to connect the tetrahedral layer in SCO and the octahedral layer in STO.

The most likely scenario for the observed lattice modulation is the symmetry mismatch between STO and SCO. Symmetry mismatch between the film and substrate are known to influence at least the first atomic layer.[6,10,40] In our film, the symmetry mismatch is the result of the planar symmetry differences between the CoO$_4$ tetrahedra and TiO$_6$ octahedra that lead to asymmetric cation displacements at the interface as evidenced by the contraction and expansion of atomic spacing. Related perovskite oxides containing mixed valent *B*-site cations can exhibit modulations in the atomic spacing due to breathing distortions, in which the size of the polyhedra can change in order to accommodate charge disproportion and the resultant changes in the ionic radii.[41,42] If this is the case at the SCO-STO interface, we might expect changes in the next octahedral layer—which are not observed. The most likely scenario for the atomic displacement of tetrahedral cobalt is likely due to the fact that the oxygen sublattices of the brownmillerite tetrahedral layer and the octahedral layer of the substrate do not lie on top of one another due to the different planar symmetry. In a perovskite-perovskite interface, octahedral tilting of the first few atomic layers of the films is often adjusted to accommodate for this kind of symmetry mismatch.[7,11] Thus, it should not be unexpected that modulation of the tetrahedral layer would be required for maintaining epitaxial growth on a perovskite substrate. Coupled with the GPA analysis, these results indicate that formation of modulated tetrahedra at the interface is the mechanism in which the brownmillerite structure relieves strain applied across the vacancy channels.

In summary, we have determined the interfacial and bulk structure of symmetry-mismatched epitaxial SCO on STO through high-resolution STEM imaging. We have found that brownmillerite SCO preferentially nucleates with the oxygen-deficient, tetrahedral sublayer on $TiO_2$-terminated STO perovskite substrates. This interfacial atomic configuration is accompanied by modulation of the Co−Co atomic spacing along the opening of the oxygen vacancy channel, which is linked to symmetry-mismatch driven interfacial atomic displacements. These observations not only help elucidate the interfacial structure of layered brownmillerites on perovskites, but suggest that further control of the modified interface through well-designed superlattices could generate new physical properties or magnetic and electrical ground states not envisaged in bulk SCO.

**Experimental Section**

*Film Growth*: An epitaxial brownmillerite SCO film was deposited on an (00*l*)-oriented and nearly lattice-matched STO substrate by pulsed laser epitaxy (PLE). The brownmillerite structure was obtained by growing at a temperature of 750 °C and pressure of 100 mTorr of $O_2$. Specific details of the growth conditions are reported elsewhere.[18]

*Characterization*: The excellent phase purity and high-structural quality of the films were confirmed using a Panalytical X'Pert Pro x-ray diffractometer. No perovskite phase impurities were detected in the XRD spectra. Annular dark-field (ADF) and annular bright field (ABF) images were collected using a Nion UltraSTEM200 operated at 200 keV in order to accurately image the interfacial microstructure. A commercial program Digital Micrograph (Version 2.11, Gatan Inc. U.S.), and a free DM plug-in (FRWRtools) were used for GPA. The images were further quantified by measuring atomic positions directly from images based on a recent approach developed for STEM.[43,44]

**Acknowledgements**
The authors thank Qian He for helpful discussions on quantitative analysis of HAADF images. This work was supported by the U.S. Department of Energy, Office of Science, Basic Energy Sciences, Materials Sciences and Engineering Division.

**Figures**

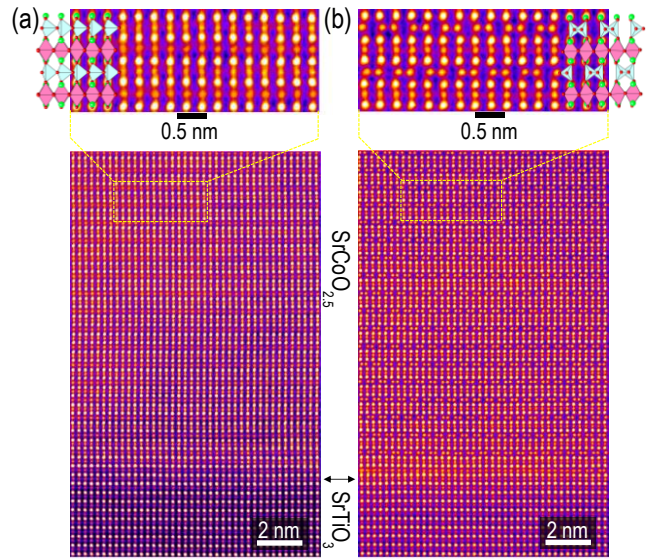

**Figure 1.** Cross-sectional ADF-STEM images seen along a) the [100] direction ($a_\parallel$ domain) and b) the [010] direction ($b_\parallel$ domain) of orthorhombic SCO, which are parallel to the [110] direction of the STO substrate. The enlarged images show a closer view of each orientation and include a structural representation of the *Ima*2 space group. Note, the two domains were observed in the same image.

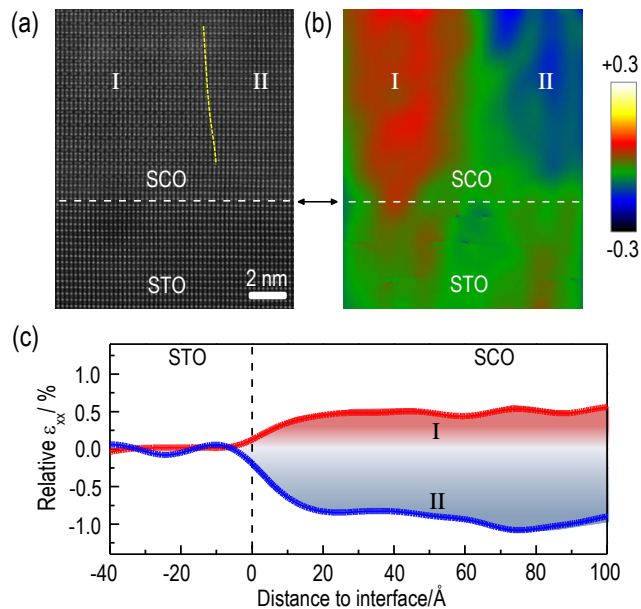

**Figure 2.** a) ADF-STEM image of SCO on STO containing both $a_\parallel$ and $b_\parallel$ domains. b) GPA $\varepsilon_{xx}$ (in-plane lattice strain) map of the image shown in (a). c) The relative strain profiles of each domain as a function of distance to the interface obtained by an intensity scan from (b). A width of 300 pixels was adopted for the intensity scan. The yellow line in (a) indicates the domain wall.

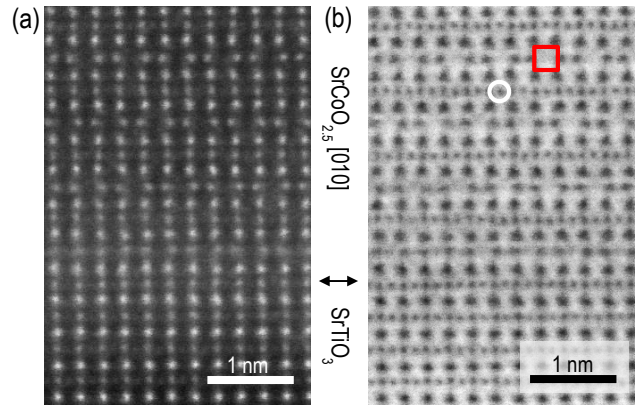

**Figure 3**. Interfacial atomic structure of a b∥ SCO domain is shown by a) Z-contrast STEM image and b) ABF image of the same film. The red square in (b) indicates the oxygen vacancy channel while the white circle indicates the oxygen position of the octahedral layer.

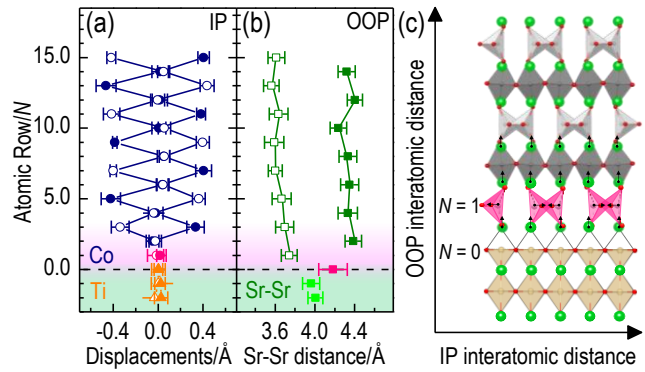

**Figure 4**. Cation interatomic spacing for a) in-plane (IP) Co and Ti atomic displacements and b) out-of-plane (OOP) Sr-Sr interatomic spacing as a function of increasing atomic layer, $N$, of the $b_\parallel$ domain. c) Schematic of orthorhombic brownmillerite and cubic perovskite interface illustrating a modified interfacial structure. Open and closed circles indicate the displacement or spacing between atomic columns reflecting lattice spacing with the same $N$. The interfacial region are highlighted in pink. The substrate interatomic spacing is for the regions when $N \leq 0$, highlighted in green.